\documentclass[runningheads]{llncs}

\usepackage{bm}
\usepackage{graphicx}
\usepackage{multirow}
\usepackage{diagbox}
\usepackage{xcolor}

\newcommand*\samethanks[1][\value{footnote}]{\footnotemark[#1]}

\begin{document}
\sloppy

\title{Utilizing Out-Domain Datasets to Enhance Multi-Task Citation Analysis}


\author{Dominique Mercier\thanks{Equal Contribution}\inst{1,2}\orcidID{0000-0001-8817-2744} \and
Syed Tahseen Raza Rizvi\samethanks\inst{1,2}\orcidID{0000-0002-4359-4772} \and
Vikas Rajashekar\samethanks\inst{1}\orcidID{0000-0002-3664-5156} \and
Sheraz Ahmed\inst{1}\orcidID{0000-0002-4239-6520} \and
Andreas Dengel\inst{1,2}\orcidID{0000-0002-6100-8255}
}

\authorrunning{D. Mercier et al.}

\institute{German Research Center for Artificial Intelligence (DFKI) GmbH\\
Trippstadter Straße 122, 67663 Kaiserslautern, Germany. \\
EMail: \email{first\_name.last\_name}@dfki.de \and
TU Kaiserslautern\\
Erwin-Schrödinger-Straße 52, 67663 Kaiserslautern, Germany.
}

\maketitle

\begin{abstract}
Citations are generally analyzed using only quantitative measures while excluding qualitative aspects such as sentiment and intent. However, qualitative aspects provide deeper insights into the impact of a scientific research artifact and make it possible to focus on relevant literature free from bias associated with quantitative aspects. Therefore, it is possible to rank and categorize papers based on their sentiment and intent. For this purpose, larger citation sentiment datasets are required. However, from a time and cost perspective, curating a large citation sentiment dataset is a challenging task. Particularly, citation sentiment analysis suffers from both data scarcity and tremendous costs for dataset annotation. To overcome the bottleneck of data scarcity in the citation analysis domain we explore the impact of out-domain data during training to enhance the model performance. Our results emphasize the use of different scheduling methods based on the use case. We empirically found that a model trained using sequential data scheduling is more suitable for domain-specific usecases. Conversely, shuffled data feeding achieves better performance on a cross-domain task. Based on our findings, we propose an end-to-end trainable multi-task model that covers the sentiment and intent analysis that utilizes out-domain datasets to overcome the data scarcity.


\end{abstract}

\section{Introduction}
Neural Networks have recently been applied to tasks from a wide range of domains. They are also notorious for their desire for very large amounts of annotated data, one of the key requirements to use neural networks is the availability of annotated data. While the process of data annotation can be automated in some domains to ensure the availability of the necessary data. However, it is not always possible and the quality of automatically annotated data can not be ensured as mentioned in~\cite{snow2008cheap}.

Citations data for sentiment analysis is a particular example of such a scenario where the data is already very scarce and challenging to collect and annotate using automated approaches. While the annotation of product reviews can be automated, the automated annotation of texts without additional features like stars ratings and emojis, etc is a significantly more complex task~\cite{ranjan2017automatic}. 

Scientific publications play an important role in the progress of a community. The "Publish or Perish" principle continuously pushes the researchers to periodically publish their scientific contributions which resulted in a boom of publications. This exponential increase in the amount of existing scientific publications has posed a challenge of evaluating the impact of each contribution in this publication outburst. Despite the existence of various metrics, including the h-index, aspects such as sentiment and intent are rarely evaluated. It is a well-established fact that most of the existing metrics heavily rely on citation counts and therefore only take quantitative aspects of a citation into consideration~\cite{bornmann2007we}. However, the quality of a scientific contribution should not entirely depend on quantitative aspects rather on the content and the results~\cite{garfield1979citation,yousif2019survey}.

Such qualitative facet greatly assists in the citation impact measurements by enriching them and therefore resulting in more sophisticated significance rankings~\cite{yousif2019survey}. The task of sentiment classification offers contextual insights into a given text corpus and is applied on various domains such as movie review, product reviews, and Twitter data~\cite{bahrainian2013sentiment,wu2015sentiment,feldman2013techniques,lin2009joint,medhat2014sentiment}. Performing sentiment analysis on objective citation data is still challenging due to the objectivity of the text and the limited amount of annotated data.

The intent of a citation found in scientific literature refers to the purpose of citing the existing scientific artifacts. Citation intent analysis serves a dual purpose. Besides the intention of a citation i.e. approach, dataset, survey, or related work, it also plays a crucial role in identifying the sentiment~\cite{mercier2019senticite} of that citation based on its occurrence position in the paper. For instance, citations found in the evaluation and discussion section are more likely to be negative, as the citing authors usually compare the results of their approach in evaluation to prove the superiority of their proposed approach. 

Despite the recently published approaches~\cite{beltagy2019scibert} there is still a scarcity of methods and datasets for the task of scientific citation analysis. There are a couple of factors that caused this data scarcity. Firstly, the high costs of manual annotation and the highly objective text make it impossible to automatically annotate it with a high quality. Secondly, there is no formal definition of intention used to classify citations properly.

In our previous paper ImpactCite~\cite{mercier2020impactcite}, we contributed by releasing a cleaned citation sentiment dataset for the task of citation sentiment analysis. In addition, we proposed a transformer-based approach for classifying the sentiment or intent of a given citation string. Even with our dataset contribution, the scarcity of citation sentiment data was not eliminated. Therefore, in this paper, we further investigated the usage of out-domain sentiment datasets to learn and transfer their knowledge to the citation sentiment task. For this purpose, we utilize the cleaned dataset proposed in ImpactCite~\cite{mercier2020impactcite} and extend its analysis using out-domain data to further improve the performance of the citation analysis model. We also evaluate different scheduling methods to train models on the data and investigate the impact of those strategies concerning the model performance. Furthermore, we investigate the impact of training a single model for two different tasks of the same domain to enhance the accuracy of the task with limited annotated data. It will significantly save both time and computation resources. To our knowledge, it is the first endeavor of investigating sentiment and intent classification on scientific data including out-domain data integration. Citation sentiment analysis would benefit greatly if sentiment datasets from other domains had a positive effect on citation analysis models. The contributions of this publication are as follows:
\begin{enumerate}
    \item Evaluation of out-domain data usage during training
    \item Evaluation of different scheduling methods
    \item An end-to-end sentiment and intent citation classification multi-task model
\end{enumerate}

\section{Related Work}
This section discusses the literature related to three relevant aspects. Firstly, we will discuss the works related to the sentiment classification followed by the intent classification. Later, we cover literature related to the use of out-domain data, transfer learning, and the possible impact of these approaches concerning data scarcity.

\subsection{Sentiment Classification}
Sentiment analysis has been a notable task in natural language processing. Several approaches have been proposed in the existing literature which focused on tackling the task of Sentiment classification. The most common use cases for the task of sentiment classification include the sentiment analysis of tweets, movies, and products reviews. 
Tang et al.~\cite{tang2014learning} proposed a word embeddings-based approach contingent on sentiment present in a tweet. These sentiment-oriented word embeddings make this approach very suitable for the task of sentiment classification. Thongtan et al.~\cite{thongtan-phienthrakul-2019-sentiment} took it one step further and applied document embeddings instead of word embeddings. These document embeddings were trained using cosine similarity as the similarity measure. The effectiveness of this approach was demonstrated by applying it to a dataset consisting of movie reviews. Cliche~\cite{cliche-2017-bb} adopted a slightly different path and employed an ensemble of Convolutional Neural Network (CNN) and Long-Short Term Memory (LSTM) models. This approach was trained and fine-tuned on a large corpus of unlabeled tweets for the task of sentiment classification.

BERT~\cite{devlin2018bert} is considered as the most popular choice for different natural language processing (NLP) tasks. It was trained on a large corpus of unlabeled data. Owing to its success in resolving other NLP problems, BERT has also been applied to the task of sentiment analysis. Several approaches~\cite{8947435,zhou2016attention,DBLP:journals/corr/abs-1904-12848} took advantage of the baseline BERT model and further tapped the potential of the model by incorporating different modules like pre-processing, attention, and structural features, etc. These modules provided some additional information to the model which in turn helped the model to better predict the resultant label.

Most of the research related to the task of sentiment analysis is performed for the domains of movie/product review or Twitter data sentiment analysis. However, a minor fraction of the literature also targets a different domain for the task of sentiment analysis. Citation sentiment analysis is also of extreme importance as it helps us in understanding the impact of research artifacts in a scientific community. Citation sentiment analysis is vastly different from movie/product review or Twitter sentiment analysis, unlike reviews and tweets, citations appear in the scientific literature which is a quite formal form of text. Esuli and Sebastiani~\cite{esuli2006determining} proposed the idea that sentiment classification has striking similarities with opinion and subjectivity mining. 
They further discussed that an individual can premeditate a seemingly positive or negative citation by only using their inclinations and writing style.

Athar et al.~\cite{athar:2011:SS} explored the idea of using sets of several features like science lexicon, contextual polarity, dependencies, negation, sentence splitting, and word-level features for citation sentiment classification. They performed several experiments to establish a set of most suitable features which has optimal performance in classifying citation sentiment found in scientific literature. On a similar line, Xu et al.~\cite{xu2015citation} carried out a citation sentiment analysis on the clinical trial literature. For this task, they employed a different set of features like n-grams, sentiment lexicon, and structure information. The task of sentiment classification is particularly hard for citation data due to the inadequate number of datasets available which have a very limited number of samples for sufficiently training a model. Finding a sentiment in a text that is written to be analytical and objective is substantially different from doing so in highly subjective text pieces like Twitter data.

\subsection{Intent Classification}
Intent classification and sentiment classification seem to be nearly identical tasks. However, both tasks are inherently different as intent classification is more inclined towards motive behind citation which is generally closely related to the section in which the citation string appears. Intent classification has become a more challenging task due to the increasing usage of compound section titles. Cohan et al.~\cite{cohan2019structural} employed bi-directional LSTM equipped with an attention mechanism. Additionally, they proposed to use ELMo vectors and structural scaffolds i.e. citation worthiness and section title.

Another interesting work is SciBERT which is a variation of BERT specifically optimized for scientific publications and was proposed by Beltagy et al.~\cite{beltagy2019scibert}. The model was trained on 1.14 million scientific publications containing $3.17$ billion tokens. The training data originates from two different domains, namely the computer science and biomedical domain. SciBERT was successfully applied on several NLP tasks including the classification of sections.

Furthermore, Mercier et al.~\cite{mercier2019senticite} tackled the sentiment and intent classification using a fusion approach of different baseline classifiers such as a Support Vector Machine (SVM) and a perceptron. They used a set of textual features consisting of adjectives, hypernyms, type, length of tokens, capitalization, and synonyms. 
Closely related to that, Abu-Jabra et al.~\cite{abu-jbara-etal-2013-purpose} proposed an SVM-based approach to perform the intent classification of citations. They stated that structural and lexical features in their experiments have shown to be of very high significance when it comes to the intent of a citation.

\subsection{Out-Domain Data Utilization}
Su et al.~\cite{su2019generalizing} presented in their work to study the impact of out-domain data for question answering. They investigated different training schedules and their impact on accuracy.
The main focus of their work was a better generalization. Another work that conducted experiments related to the robust training using in-domain and out-domain data was proposed by Li et al.~\cite{li2018s}. Their proposed method provides the capabilities to learn domain-specific and general data in conjunction to overcome the convergence towards domain-specific properties. Sajjad et al~\cite{sajjad2017neural} proposed an approach that first learns of different out-domain data and finally fine-tunes on in-domain data to achieve the optimal results. This approach intuitively utilizes the data of the different domains and therefore has a much larger training corpus for a better generalization.

Khayrallah et al.~\cite{khayrallah2018regularized} addressed the amount of out-domain vocabulary. Their findings showed that with the use of out-domain data and a continuous adaption towards the domain, the number of words not included in the vocabulary can be reduced efficiently. For this purpose, they used an out-domain model and trained it with a modified training objective continuously on the in-domain data. Furthermore, Mrk{\v{s}}i{\'c} et al.~\cite{mrkvsic2015multi} showed that using the out-domain data can yield significant improvements for very small datasets. And therefore makes it possible to train models using these sets when it is not possible to do that without the use of out-domain data.

\section{Datasets}
This paper mainly focuses on the task of sentiment and intent analysis. Therefore we selected a range of datasets suitable for sentiment classification and also for intent classification.

\subsection{Sentiment Datasets}
For the task of Sentiment classification, we employed various datasets for our experiments. Our target domain is the scientific literature. However, we selected some out-domain datasets to overcome the data scarcity. Following are the datasets selected for the sentiment classification task:

\begin{enumerate}
    \item Movie reviews
    \item Product reviews
    \item Twitter data
    \item Scientific data
\end{enumerate}

To standardize the labels of selected datasets, a preprocessing step was essential. For experiments evaluating out-domain knowledge transfer and sequential training, we preprocessed the selected datasets for binary sentiment classification tasks i.e. positive and negative. It enabled us to train and test models across different datasets. To do so, we excluded the neutral class and grouped different labels if the datasets had multiple classes that correspond to the positive or negative label e.g. 'good' and 'very good' or $4$ out of $5$ and $5$ out of $5$ stars. However, we used all three classes i.e. positive, negative, and neutral for the multi-task experiments. The details of the selected sentiment datasets are as follows:

\subsubsection{Movie Reviews:}
From the domain of movie reviews, we decided to use three popular datasets that quantified both positive and negative reviews in the form of a numerical score. The IMDB~\cite{maas-EtAl:2011:ACL-HLT2011} dataset contains about $25,000$ training and $25,000$ test instances of highly polar reviews. It is the largest dataset by volume in the selected datasets. The second dataset we used in our experiments is the Cornell movie review data~\cite{Pang+Lee:04a}. It is a considerably small dataset as compared to IMDB. However, it has an even distribution of $1,000$ samples for each of the positive and negative classes. The last dataset that we selected from movie reviews is the Stanford Sentiment Treebank dataset~\cite{socher2013recursive}. For this dataset, we had to discard the samples not related to either negative or positive classes. All three above-mentioned datasets are related to the same task from the same domain and therefore their underlying structure should be rather similar.

\subsubsection{Product Reviews:}
To include a dataset from a different domain than the Movie reviews, we selected the amazon product review dataset~\cite{mcauley2015image}. This dataset consists of various product categories. Some of the categories in the amazon data are closely related to the movie reviews such as Books, TV, and Movies. On the other hand, some categories are completely different from movie reviews such as Beauty, Electronic, and Video Games. For our experiments, we selected one category from amazon data that was unrelated to the movie reviews. The chosen category was related to the instrument reviews. The product reviews were quantified in the form of $1-5$ stars. For our experiments, we converted the star ratings into positive and negative classes while skipping the neutral class. Product reviews with ratings with $4$ and $5$ stars were labeled as positive. On the other hand, product reviews with $1$ and $2$ stars were labeled as negative. However, product reviews with a star rating of $3$ were skipped as they belonged to the neutral class and were not relevant for our experiments.

\subsubsection{Twitter Data:}
Sentiment analysis on Twitter data is a quite popular task. For this purpose, we selected a couple of Twitter datasets. Intuitively, we assume that the Twitter datasets are the most subjective ones in our selection as their language style differs significantly from the scientific and other domain datasets. The first dataset is related to airline reviews in form of tweets. The data was taken from Kaggle~\footnote{Twitter US Airline Sentiment:~\url{https://www.kaggle.com/crowdflower/twitter-airline-sentiment}} and contains three classes i.e. positive, negative, and neutral. Similar to other datasets, we removed the neutral class. The same class elimination was performed for the second dataset Sentiment140 dataset~\footnote{Sentiment140:~\url{https://www.kaggle.com/kazanova/sentiment140}}. This dataset was composed using $1.6$ Million general tweets collected from Twitter along with their sentiment.

\subsubsection{Scientific Data:}
From the scientific domain, we selected a dataset called Citation Sentiment Corpus (CSC-Clean). It was proposed in our previous paper~\cite{mercier2020impactcite}. There have been very limited contributions for the citation sentiment analysis task as the number of available datasets is almost not existent. Although there exist some datasets proposed by Xu et al.~\cite{xu2015citation} and Athar~\cite{athar:2011:SS} those are either not publicly available or suffer from bad quality. The reason for this data scarcity is the expensive and complicated labeling process. We decided to use the CSC dataset~\cite{athar:2011:SS} as a baseline. Upon careful dataset analysis, we found out that there exist duplicate instances with an occasionally same or different label in the CSC dataset. These instances also often exist in different data splits such as training and test set. We identified these quality issues and cleaned the dataset to achieve a better quality throughout the same corpus. Table~\ref{tab:csc_comparison} shows the original sample count, number of removed instances concerning duplicates, and the remaining number of samples. In addition, we show the updated dataset distribution and the percentage of removed instances concerning each class. In total, we removed $757$ instances which are $8.67\%$ of the data. For duplicates with two different labels, we removed both the original and the duplicated instances as this is the only appropriate solution to avoid a subjective bias from our side. Including one of the instances would bias the data and results. The resultant dataset is referred to as CSC-Clean and publicly available~\footnote{~\url{https://github.com/DominiqueMercier/ImpactCite}}.

\begin{table}[!t]
\renewcommand{\arraystretch}{1.3}
\caption{Comparison of citation sentiment corpus (CSC) and citation sentiment clean (CSC-C) dataset. Taken from~\cite{mercier2020impactcite}.}
\label{tab:csc_comparison}
\centering
\begin{tabular}{l|c|c|c|c}
\textbf{Classes} & \textbf{CSC} & \textbf{CSC-Clean} & \textbf{CSC-Clean Dist.} & \textbf{Removed [\%]} \\
\hline
Positive    & $829$     & $728$     & $9.12\%$  & $101$ ($12.18$) \\
\hline
Neutral     & $280$     & $253$     & $87.71\%$ & $27$  ($9.64$) \\
\hline
Negative    & $7,627$   & $6,999$   & $3.17\%$  & $629$ ($8.25$) \\
\end{tabular}
\end{table}

\subsubsection{Sentiment Dataset Statistics:}
In Table~\ref{tab:datasets} we show the statistics of each sentiment dataset after pre-processing them to exclude the neutral class and existing duplicates. These statistics include the number of samples used to train, validate, and test our models. In addition, the table also shows the dataset distribution highlighting that datasets such as the Instruments, US Airline, and CSC-Clean are heavily biased towards one of the two classes. Another characteristic is that the collected datasets differ largely in their size. This resulted in the need to upsample or downsample the data for some experiments to make the results comparable.

\begin{table}[!t]
\renewcommand{\arraystretch}{1.3}
\caption{Comparison all used datasets. Only including the positive and negative class. Neutral class for CSC-Clean was excluded in this table.}
\label{tab:datasets}
\centering
\begin{tabular}{l|l||c|c|c||c|c}
\textbf{Domain} & \textbf{Dataset} & \textbf{Train} & \textbf{Val} & \textbf{Test} & \textbf{Positive [\%]} & \textbf{Negative [\%]} \\
\hline
\multirow{3}{*}{Movie Reviews} 
& IMDB              & $19,923$  & $4,981$   & $24,678$  & $50.19$   & $49.81$ \\
& Cornell           & $6,823$   & $1,706$   & $2,133$   & $50.0$    & $50.0$ \\
& Stanford Sent.    & $6,911$   & $872$     & $1,819$   & $51.64$   & $48.36$ \\
\hline
Product Reviews 
& Instruments       & $6,068$   & $1,507$   & $1,897$   & $95.07$   & $4.93$ \\
\hline
\multirow{2}{*}{Twitter Data} 
& US Airline        & $7,243$   & $1,811$   & $2,264$   & $19.81$   & $80.19$ \\
& Sentiment140      & $10,161$  & $2,541$   & $3,176$   & $49.94$   & $50.06$ \\
\hline
Scientific Data
& CSC-Clean         & $797$     & $89$      & $95$      & $74.21$   & $25.79$ \\
\end{tabular}
\end{table}

\subsection{Intent Dataset}

\begin{table}[!t]
\renewcommand{\arraystretch}{1.3}
\caption{SciCite~\cite{cohan2019structural}. Number of instances and class distribution. Taken from~\cite{mercier2020impactcite}.}
\label{tab:scicite_data}
\centering
\begin{tabular}{l|c|c|c|c|c}
\textbf{Classes} & \textbf{Training} & \textbf{Validation} & \textbf{Test} & \textbf{Total} & \textbf{Percentage} \\
\hline
Result              & $1,109$    & $123$ & $259$ & $1,491$    & $13.53$ \\
\hline
Method              & $2,294$    & $255$ & $605$ & $3,154$    & $28.62$ \\
\hline
Background          & $4,840$    & $538$ & $997$ & $6,375$    & $57.85$ \\
\end{tabular}
\end{table}

From the scientific domain, we selected a dataset related to citation intent analysis called SciCite. The SciCite dataset proposed in~\cite{cohan2019structural} is a famous benchmark for citation intent classification. It was curated using medical and computer science publications and is publicly available. The size of this dataset is sufficient to train any deep learning model and the existing benchmarks emphasize the high quality of the dataset. However, the dataset has an imbalanced sample distribution in which the vast majority of the samples are assigned to the 'Background' class. Another, important aspect of the dataset is the coarse-grained label process which was applied to create that dataset. According to the authors, the distribution follows the real-world distribution and the number of samples is large enough to sufficiently learn the concepts of each class. Detailed information about the dataset can be found in Table~\ref{tab:scicite_data}. We mainly employed SciCite along with the CSC-Clean dataset to demonstrate the capability of training a multi-task model, where tasks are different and yet from the same domain.

\section{Contributions}
We divided this section into three main parts. The first part discusses the baseline work from our previous paper ImpactCite~\cite{mercier2020impactcite}. Secondly, we will discuss the impact of training a model on out-domain data. and the third part covers a fusion approach to combine sentiment and intent. We further show that both methods rely on different aspects of the task and highlight their advantages.

\subsection{ImpactCite}
Our previously proposed approach, ImpactCite~\cite{mercier2020impactcite} served as a baseline for this paper. It is an XLNet~\cite{yang2019xlnet} based approach for analysis of sentiment and intent of citations found in scientific literature. ImpactCite utilizes two separate XLNets to provide a citation sentiment and intent analysis. To the best of our knowledge, there exists only limited work concerning scientific citation analysis. 

\begin{figure}[!t]
\centering
\includegraphics[width=0.5\linewidth]{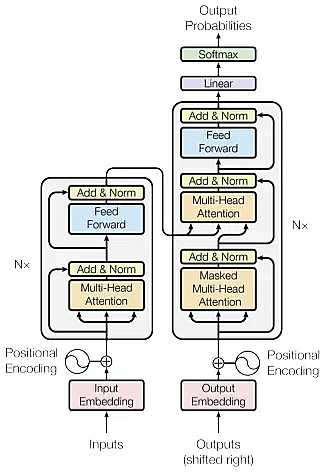}
\caption{Transformer-XL architecture~\cite{dai2019transformer}. Each of the Multi-Head Attention layers is composed of multiple attention heads that apply a linear transformation and compute the attention.}
\label{fig:transformer_xl}
\end{figure}

The task of citation analysis involves two challenging dataset characteristics. First, the dependency on sentences next to the actual citation. Taking into account that most of the citation sentiment origins from neighborhood sentences lead to longer sequences. Secondly, the model needs to cover dependencies in both directions as in the scientific world the sentiment might be given before or even after the actual citation sentence. Taking into account these essential properties we decided to use XLNet~\cite{yang2019xlnet} as a model for our experiments. XLNet is a well-known transformer-based network structure that can cover long sequences and bi-directional dependencies. The auto-regressive model is based on a Transformer-XL~\cite{dai2019transformer} as the backbone. The Transformer XL architecture is shown in Figure~\ref{fig:transformer_xl}. In addition, there exist many pre-trained XLNet models which is essential for the sentiment classification as the number of datasets for scientific sentiment citation is not sufficient to train such a model from scratch. Precisely, we decided to use the XLNet-Large model to make sure that the model is large enough to cover the whole context. XLNet-Large consists of $24$-layers, $1,024$ hidden units, and $16$ heads. During our experiments, we only fine-tune the pre-trained model according to the different tasks involving cross-domain sentiment analysis, scientific sentiment classification, and scientific intent classification. As the language of the pre-trained model and the data used to fine-tune it we benefit from the pre-trained weights as the general language structure is similar and only needs small adjustments concerning the domain and task.

Separating these two tasks enables us to fine-tune the corresponding model to each task and achieve the best possible results for that task. This is especially beneficial for the intent as the amount of sentiment citation data is limited. However, the major drawback is that two separate models are required for this purpose and the sentiment does not benefit from the intent model although both tasks are from the same domain.

\subsection{Overcoming Data Scarcity \& Data Feeding Techniques}
In this paper, we investigated the techniques to overcome the scarcity of data for certain domains. Particularly for sentiment analysis of scientific citations, there are not many datasets available. In this paper, we propose that training on out-domain data and later finetuning on target domain results in better model performance, therefore, bridging the data scarcity gap. Additionally, we experimented with different data feeding methods to analyze their impact on the performance of the final model.

\subsection{Fusion Approach}
\begin{figure}[!t]
\centering
\includegraphics[width=0.6\linewidth]{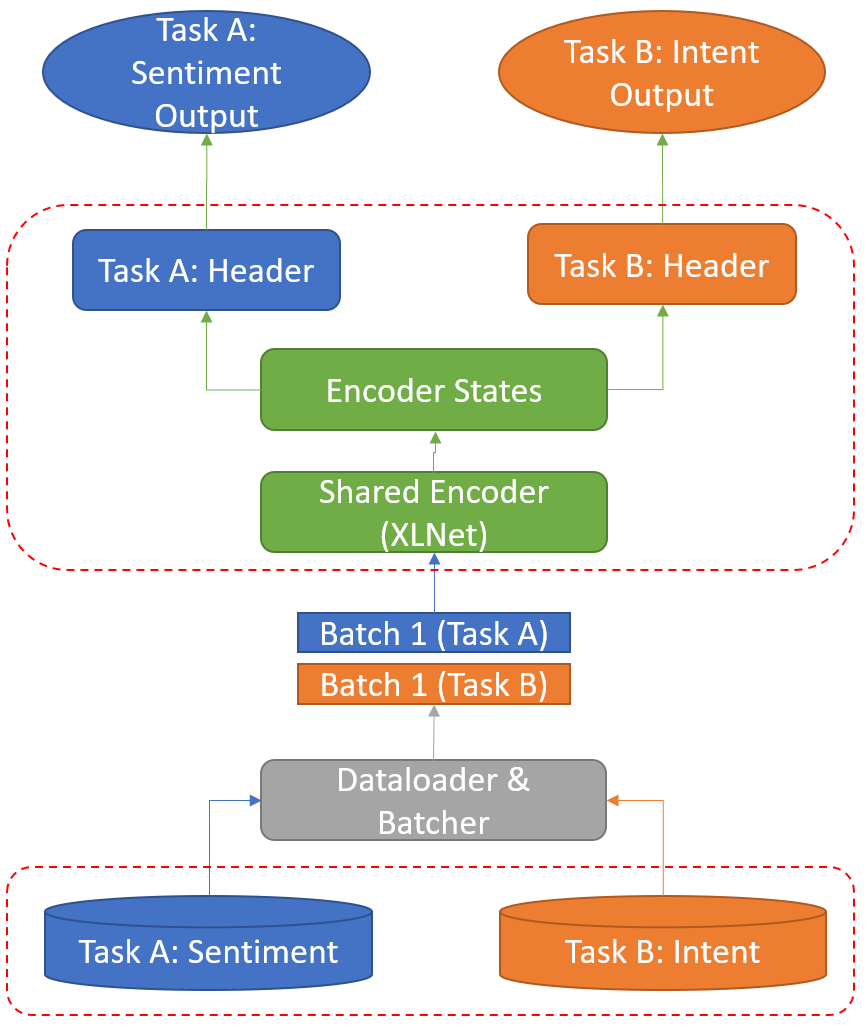}
\caption{Mutli-Task setup combining sentiment and intent task. The same encoder is used for both tasks and a task specific head is trained.}
\label{fig:mutli-task_setup}
\end{figure}

Lastly, in this paper, we propose that although the citation sentiment and intent analysis are different tasks. However, we believe that the underlying text structure concerning the sentiment and intent task on scientific data is similar. Based on the cross-domain sentiment classification we show that the addition of data addressing the same task or the same domain can enhance the scientific sentiment classification. Ultimately, we train a single XLNet model on both the sentiment and intent datasets that performs the complete citation analysis and resolves the dataset size issues. The pipeline is visualized in Figure~\ref{fig:mutli-task_setup}.

\section{Experiments and Analysis}
In this section, we will discuss our experiments and their results. All experiments are classified into four sets. The first set discusses the performance benchmark of XLNet~\cite{yang2019xlnet} for the task of intent classification. The second set discusses the experiments related to the performance benchmark of XLNet for sentiment classification. We performed these benchmarking experiments using several other models ranging from the baseline models i.e. CNN to highly sophisticated language models i.e. BERT~\cite{devlin2018bert}, ALBERT~\cite{lan2019albert} and XLNet~\cite{yang2019xlnet}. In the third set of experiments, we will discuss the experiments related to training on out-domain data and testing on several different domains dataset, which also includes finetuning on the target domain dataset. Additionally, a collection of experiments discussing the effects of different data feeding techniques are also discussed in this set of experiments.

Finally, we discuss experiments combining the sentiment and intent modality and serve a single model that processes both tasks. Doing so requires a deep understanding of multiple aspects such as domain dependency, model selection, and task relation. In our case, the first two aspects are covered by benchmarking and the out-domain evaluation. In addition, it has been shown that the tasks are related to each other~\cite{mercier2019senticite}.

\subsection{Intent Classification}
\subsubsection{Experiment1: Performance Benchmarking.}
To evaluate the performance of different model architectures on the intent classification task we decided to use the SciCite dataset~\cite{cohan2019structural}. We used the original train and test splits provided by the dataset and divided our models into two categories. The first category includes all baseline models. We explored different setups of CNNs, LSTMs, and RNNs. These models were trained from scratch using the SciCite dataset. In addition, we also trained BERT~\cite{devlin2018bert}, ALBERT~\cite{lan2019albert} and ImpactCite (XLNet). The second category of models was pre-trained and is a member of the transformer-based solutions. These models were only fine-tuned on the SciCite dataset. Due to high imbalance data, we employed the micro-f1 and macro-f1 scores for performance comparison. Furthermore, initial experiments using the CNN, LSTM, and RNN approaches have shown that their performance using pre-trained embeddings e.g. {GloVe}\footnote{https://nlp.stanford.edu/projects/glove/} did not improve compared to newly initialized embeddings. We emphasize that one of the reasons for this might be the domain discrepancy between the pre-trained embeddings and the scientific domain.

\subsubsection{Results and Discussion:}

\begin{table*}[!t]
\renewcommand{\arraystretch}{1.3}
\caption{Performance evaluation on SciCite~\cite{cohan2019structural} (intent) dataset. L = Layer, F = Filter, C = convolution size. Taken from~\cite{mercier2020impactcite}.}
\label{tab:scicite_eval}
\centering
\scalebox{0.8}{
\begin{tabular}{l|l||c|c|c||c|c}
\multirow{2}{*}{\textbf{Topography}} & \multirow{2}{*}{\textbf{Architecture}} & \multicolumn{3}{c||}{\textbf{Class-based accuracy}} & \multirow{2}{*}{\textbf{micro-f1}} & \multirow{2}{*}{\textbf{macro-f1}} \\
\cline{3-5}
& & \textbf{Result [\%]} & \textbf{Method [\%]} & \textbf{Background [\%]} &  &  \\
\hline\hline
CNN & L 3 F 100 C 3,4,5 & $79.92$ & $76.53$ & $79.24$ & $78.50$ & $78.56$ \\
\hline
CNN & L 3 F 100 C 2,4,6 & $81.85$ & $77.69$ & $81.14$ & $80.12$ & $\bm{80.22}$ \\
\hline
CNN & L 3 F 100 C 3,3,3 & $64.09$ & $71.74$ & $85.46$ & $78.05$ & $73.76$ \\
\hline
CNN & L 3 F 100 C 3,5,7 & $76.45$ & $74.05$ & $85.46$ & $\bm{80.49}$ & $78.65$ \\
\hline
CNN & L 3 F 100 C 3,7,9 & $68.34$ & $70.58$ & $87.26$ & $79.20$ & $75.39$ \\
\hline
LSTM & L 2 F 512        & $73.75$ & $73.55$ & $79.54$ & $76.80$ & $75.61$ \\
\hline
LSTM & L 4 F 512        & $75.29$ & $69.59$ & $82.95$ & $77.54$ & $75.94$ \\
\hline
LSTM & L 4 F 1024       & $68.73$ & $70.91$ & $84.25$ & $77.75$ & $74.63$ \\
\hline
RNN & L 2 F 512         & $25.10$ & $56.86$ & $62.19$ & $55.30$ & $48.05$ \\
\hline
\hline
BERT~\cite{devlin2018bert} & Base 
                        & $84.56$ & $75.37$ & $89.47$ & $84.20$ & $83.13$ \\
\hline
ALBERT~\cite{lan2019albert} & Base 
                        & $83.78$ & $77.03$ & $87.06$ & $83.34$ & $82.62$ \\
\hline
ImpactCite~\cite{mercier2020impactcite} & Base 
                        & $92.67$ & $85.79$ & $88.34$ & $\bm{88.13}$ & $\bm{88.93}$ \\
\hline
\hline
BiLSTM-Att~\cite{cohan2019structural} & *               & * & * & * & * & $82.60$ \\
\hline
Scaffolds~\cite{cohan2019structural} & *                & * & * & * & * & $84.00$ \\
\hline
BERT~\cite{beltagy2019scibert,devlin2018bert} & Base    & * & * & * & * & $84.85$ \\
\hline
SciBert~\cite{beltagy2019scibert} & *                   & * & * & * & * & $85.49$ \\
\end{tabular}
}
\end{table*}

Table~\ref{tab:scicite_eval} shows the performance benchmark results of different selected architectures for the intent classification task. It is evident from the results that both the LSTM and RNN are not able to compete with the CNN. A reason for the inferior performance of the RNN is the length of the sequences resulting in vanishing gradients for the RNN. The LSTM on the other hand suffers from the bi-directional influences between the sentences that are not completely covered by the architecture. We further explored different layer and filter sizes for baseline models. However, there is only an insignificant difference when tuning the parameters. Concerning the time consumption, the CNN shows superior performance over the other baseline approaches as it can compute things in parallel as compared to LSTMs and RNNs.

The second category presented in Table~\ref{tab:scicite_eval} shows the complex language models. We were able to achieve a new state-of-the-art performance using ImpactCite~\cite{mercier2020impactcite}. It significantly outperformed the other fine-tuned language models by up to $3.9\%$ micro-f1 and $5.8\%$ macro-f1 score. Especially, the increase in the minority classes has shown a significant difference of $10\%$. Summarizing the findings, we have demonstrated that ImpactCite (XLNet) was able to outperform the CNN by $8.71\%$ and the language models by $3.9\%$ macro-f1 score and significantly increased the performance for the minority class. This highlights the significantly better capabilities of the larger transformer-based model pre-trained on a different domain and later fine-tuned.

\subsection{Sentiment Classification}
In this section, we will discuss the experiments conducted for the task of scientific sentiment classification. There were two datasets used in these experiments namely Citation Sentiment Corpus (CSC) and our proposed clean version of the dataset called CSC-C.

\subsubsection{Experiment 1: Fixed Dataset Split on CSC Sentiment Dataset.}
For this experiment, we employed a fixed $70$/$30$ data split for the CSC dataset excluding any additional dataset cleansing. We evaluated the performance of each previously used model. Additionally, we employed several sample strategies i.e. focal loss, SMOTE \& upsampling, and analyzed their impact concerning the imbalanced data.

\subsubsection{Results and Discussion:}
The results of this experiment are shown in Table~\ref{tab:sentiment_csc}. We observed that all models mainly captured the concept of neutral citations. Additionally, we also observed that the methods like focal loss and SMOTE sampling increased the performance of the CNNs and LSTMs. Furthermore, upsampling does not help to improve the performance of the model. However, ImpactCite~\cite{mercier2020impactcite} effectively learned representations of each class. Especially, the negative class was captured in a much better way by ImpactCite. Although ImpactCite showed slightly worse performance on the neutral class, it performed significantly better for positive and negative classes. We conclude that ImpactCite is able to deal with the large class imbalance and show that the complex language models are superior to the baseline approaches enhanced with sampling and focus strategies for the CSC dataset.

\begin{table*}[!t]
\renewcommand{\arraystretch}{1.3}
\caption{Performance: Citation Sentiment Corpus (CSC). Taken from~\cite{mercier2020impactcite}.}
\label{tab:sentiment_csc}
\centering
\begin{tabular}{l|l||c|c|c}
\multirow{2}{*}{\textbf{Topography}} & \multirow{2}{*}{\textbf{Modification}} & \multicolumn{3}{c}{\textbf{Class-based accuracy}} \\
\cline{3-5}
& & \textbf{Positive [\%]} & \textbf{Negative [\%]} & \textbf{Neutral [\%]} \\
\hline\hline
CNN & *                         & $28.2$ & $21.3$ & $94.8$ \\
\hline
CNN & Focal                     & $36.9$ & $16.9$ & $94.3$ \\
\hline
CNN & SMOTE                     & $39.4$ & $20.2$ & $84.2$ \\
\hline
CNN & Upsampling                & $36.1$ & $6.7$  & $92.8$ \\
\hline
\hline
LSTM & *                        & $32.8$ & $12.4$ & $93.9$ \\
\hline
LSTM & Focal                    & $42.7$ & $19.1$ & $82.8$ \\
\hline
LSTM & SMOTE                    & $42.3$ & $20.2$ & $83.7$ \\
\hline
LSTM & Upsampling               & $26.1$ & $11.2$ & $\bm{97.0}$ \\
\hline
\hline
RNN & *                         & $24.5$ & $21.3$ & $72.7$ \\
\hline
\hline
BERT~\cite{devlin2018bert}              & * & $38.6$ & $20.4$ & $96.4$ \\
\hline
ALBERT~\cite{lan2019albert}             & * & $44.3$ & $28.8$ & $95.8$ \\
\hline
ImpactCite~\cite{mercier2020impactcite} & * & $\bm{78.9}$ & $\bm{85.7}$ & $75.4$ \\
\end{tabular}
\end{table*}

\subsubsection{Experiment 2: Cross-Validation on CSC-Clean Sentiment Dataset.}
In order to compare our proposed ImpactCite with the results of Athar~\cite{athar:2011:SS} we used a $10$-fold-cross validation. However, due to the missing split information and the duplicates that exist in the original CSC dataset, we decided to perform the experiment on the CSC-C dataset. Although the results are not directly comparable, the approach~\cite{athar:2011:SS} is favored due to the duplicates that appear in the training and test data. For the sake of completion, we included~\cite{athar:2011:SS} as a reference. During the $10$-fold cross-validation, we used nine splits as training and one split as a test dataset for each run and averaged the results at the end. A collection of experiments were performed employing a variety of models ranging from baseline CNN models to complex BERT language models. In order to successfully apply the baseline methods, we used the class weights as they have shown superior performance in previous experiments.

\subsubsection{Results and Discussion:}
The results of this experiment are shown in Table~\ref{tab:sentiment_scc_cross}. Interestingly, the baseline models were not able to achieve comparable performance even though the class weights were employed. In order to resolve the class imbalance issue, we pre-processed the folds for the baseline approaches such that the number of positive and neutral training samples was decreased to the number of negative samples. Doing so resulted in the performances shown in the table. Additionally, we observed that the complex language models performed much better on the small dataset. They significantly outperformed the baseline methods and achieved good results across all three classes. In addition, ImpactCite˜\cite{mercier2020impactcite} outperformed all other selected models and sets a new state-of-the-art for citation sentiment classification on the CSC-Clean. For the sake of completeness, we included the SVM used by Athar evaluated on the CSC dataset.

\begin{table*}[!t]
\renewcommand{\arraystretch}{1.3}
\caption{Cross validation performance: Sentiment citation corpus (CSC-C). Taken from~\cite{mercier2020impactcite}.}
\label{tab:sentiment_scc_cross}
\centering
\begin{tabular}{l||c|c|c||c|c}
\multirow{2}{*}{\textbf{Topography}} & \multicolumn{3}{c||}{\textbf{Class-based accuracy}} & \multirow{2}{*}{\textbf{micro-f1}} & \multirow{2}{*}{\textbf{macro-f1}} \\
\cline{2-4}
& \textbf{Positive [\%]} & \textbf{Negative [\%]} & \textbf{Neutral [\%]} & & \\
\hline\hline
CNN                                     & $40.2$ & $24.9$ & $95.0$ & $88.6$ & $43.4$ \\
\hline
LSTM                                    & $34.8$ & $19.0$ & $92.1$ & $84.6$ & $46.1$ \\
\hline
RNN                                     & $20.7$ & $17.9$ & $86.0$ & $77.9$ & $41.5$ \\
\hline
\hline
BERT~\cite{devlin2018bert}              & $72.8$ & $80.2$ & $70.3$ & $74.4$ & $74.4$ \\
\hline
ALBERT~\cite{lan2019albert}             & $71.1$ & $72.5$ & $67.6$ & $70.4$ & $70.4$ \\
\hline
ImpactCite~\cite{mercier2020impactcite} & $64.6$ & $86.6$ & $82.0$ & $77.7$ & $\bm{77.7}$ \\
\hline
\hline
SVM~\cite{athar:2011:SS}\footnotemark    & *      & *      & *      & $\bm{89.9}$ & $76.4$ \\
\end{tabular}
\end{table*}

\footnotetext[7]{Trained and tested on CSC}

\subsection{Out-Domain: Evaluating Impact of Additional Data}
In this section, we present our results using out-domain data to evaluate its impact on the model performance. We investigate multiple scenarios of cross dataset training and testing on datasets from different domains. Furthermore, we conducted experiments concerning the use of multiple datasets and an optimal schedule strategy to enlarge the corpus size. We also discuss details of some experiments related to different data feeding methods.

\subsubsection{Experiment 1: Out-domain Testing.}
In this experiment, we employed a pre-trained XLNet for each dataset and fine-tune it on one dataset. Once the model is trained, we evaluated its performance across all datasets to find out which datasets are semantically closer to each other. The goal is to better understand the correlation of the dataset and to what extent it is possible to use the model trained on an out-domain dataset for the prediction of sentiment across other domains. In this experiment, we trained each model for $40$ epochs with a batch size of $24$. In addition, we also used an early stopping mechanism such that if the model converges before $40$ epochs then it will stop further training to prevent over-fitting. It has to be mentioned, that in this experiment the datasets had different sizes, as shown in Table~\ref{tab:datasets}.

\subsubsection{Results and Discussion:}

\begin{table}[!t]
\renewcommand{\arraystretch}{1.3}
\caption{Results for testing on out-domain data using XLNets trained on a single dataset. Results are macro f1-scores in percent.}
\label{tab:out-domain_testing}
\centering
\scalebox{0.92}{
\begin{tabular}{l||c|c|c||c||c|c||c}
\multirow{2}{*}{\diagbox[width=6.2em]{\textbf{Train}}{\textbf{Test}}}
& \multicolumn{3}{c||}{\textbf{Movie}} & \textbf{Product} & \multicolumn{2}{c||}{\textbf{Twitter}} & \textbf{Scientific} \\
\cline{2-8}
& IMDB & Cornell & Stanford & Instruments & Us Airline & Sentiment140 & CSC-Clean \\
\hline\hline
IMDB            & $\bm{94.38}$ & $81.58$ & $83.66$ & $70.20$ & $64.53$ & $62.72$ & $54.16$ \\
\hline
Cornell         & $92.05$ & $\bm{89.69}$ & $\bm{94.39}$ & $57.46$ & $87.69$ & $69.15$ & $60.28$ \\
\hline
Stanford        & $91.71$ & $89.49$ & $92.85$ & $63.68$ & $86.46$ & $68.89$ & $63.76$ \\
\hline\hline
Instruments     & $86.51$ & $55.53$ & $57.14$ & $\bm{82.73}$ & $52.63$ & $57.52$ & $49.71$ \\
\hline\hline
US Airline      & $56.80$ & $71.45$ & $79.47$ & $43.80$ & $\bm{92.21}$ & $68.39$ & $43.45$ \\
\hline
Sentiment140    & $79.28$ & $72.63$ & $76.95$ & $65.13$ & $77.14$ & $\bm{80.57}$ & $62.42$ \\
\hline\hline
CSC-Clean       & $85.04$ & $62.91$ & $62.79$ & $64.60$ & $63.88$ & $62.13$ & $\bm{76.67}$ \\
\end{tabular}
}
\end{table}

In Table~\ref{tab:out-domain_testing} we show the results when using a single training set and testing across all datasets. Overall the best performance was achieved using the same dataset for training and testing, the only exception is the Stanford dataset. Interestingly, the performance for the Stanford dataset is surprisingly good when the model is trained on the Cornell data. It has to be mentioned, that both datasets are from the same domain. This shows that training on more domain data without fine-tuning on a specific dataset can result in a pretty good model for that dataset which is taken from the same domain. Overall training on the Stanford dataset was not successful. In general training on a dataset of the same domain without fine-tuning the model resulted in a good performance on their own domain however it is not the case when trained on out-domain data. One reason for this is the correlation between the data within the same domain. The results further show that the correlation across domains is in general lower but in the case of the Instruments dataset, the correlation is high enough to achieve superior performance using a dataset that is more balanced from the movie review domain. This suggests that a correlation between movie reviews and instrument reviews (product reviews) exists. Intuitively, this is the case because the understanding of positive and negative in the scientific domain is fundamentally different compared to review data or tweets.

\subsubsection{Experiment 2: Sequential Training.}
In this experiment, we evaluated the impact of a sequential training scheme. The idea is that if a dataset is very small and therefore it is not possible to train only on that dataset, we enhance the data size by using additional datasets. There are two interesting aspects to using additional datasets. One it will increase the amount of data available for training and secondly, we also want to evaluate the impact of the dataset sequence in which data is fed to the network. Intuitively, the last dataset category in the training sequence should be favored with respect to the performance as the gradients are optimized on it. We performed this for a fixed sequence of datasets and categories and used several permutations of the sequence of categories to have comparable results. In addition, we performed these experiments twice, once for the upsampled datasets and once for the downsampled. The reason for this procedure is that it is important to make all datasets the same size such that they can contribute the same amount to the training. With the initial dataset sizes, this would not be the case and a few datasets would dominate the training due to their size. In the upsampled version we used $3,000$ samples whereas for the downsampling experiment we used the number of instances of the smallest dataset as a reference number. For some datasets, this means we had to select a subset of the training instances. This means we do not preserve the individual class distribution. The sequence of the datasets is shown in the corresponding results tables.

\subsubsection{Results and Discussion:}

\begin{table}[!t]
\renewcommand{\arraystretch}{1.3}
\caption{Macro f1-scores for sequential training. Sequence within the categories: [P]roduct (Instruments), [M]ovie (Cornell, IMDB Stanford Sent.), [S]cientific (CSC-Clean), [T]witter (Sentiment140, US Airline). 'Up' corresponds to the upsampled training data and 'Down' to the downsampled training data.}
\label{tab:sequential}
\centering
\scalebox{0.96}{
\begin{tabular}{l||c|c|c||c||c|c||c}
\multirow{2}{*}{\diagbox[width=4.5em]{\textbf{Train}}{\textbf{Test}}}
& \multicolumn{3}{c||}{\textbf{Movie}} & \textbf{Product} & \multicolumn{2}{c||}{\textbf{Twitter}} & \textbf{Scientific} \\
\cline{2-8}
& IMDB & Cornell & Stanford & Instruments & Us Airline & Sentiment140 & CSC-Clean \\
\hline\hline
\textbf{Up} & & & & & & & \\
S T P M & $\bm{93.05}$ & $\bm{88.51}$ & $90.87$ & $80.22$ & $89.69$ & $75.45$ & $\bm{78.18}$ \\
M S T P & $92.94$ & $86.98$ & $89.35$ & $\bm{80.25}$ & $86.97$ & $\bm{77.16}$ & $69.16$ \\
P M S T & $91.62$ & $87.81$ & $90.05$ & $74.32$ & $89.84$ & $76.25$ & $70.39$ \\
T P M S & $92.19$ & $88.19$ & $\bm{91.26}$ & $77.72$ & $\bm{90.04}$ & $76.08$ & $76.97$ \\
\hline\hline
\textbf{Down} & & & & & & & \\
S T P M & $\bm{92.38}$ & $\bm{87.29}$ & $\bm{89.98}$ & $\bm{80.45}$ & $85.93$ & $\bm{76.96}$ & $\bm{75.55}$ \\
M S T P & $92.26$ & $85.65$ & $88.27$ & $78.69$ & $88.38$ & $76.13$ & $\bm{75.55}$ \\
P M S T & $90.55$ & $85.94$ & $89.27$ & $65.93$ & $\bm{88.79}$ & $75.33$ & $66.73$ \\
T P M S & $88.95$ & $83.00$ & $86.98$ & $72.07$ & $87.11$ & $75.18$ & $67.34$ \\
\end{tabular}
}
\end{table}

In Table~\ref{tab:sequential} we present the results for sequential training. The upper part of the table covers the training results using the upsampling whereas the lower part covers the downsampling results. Our results for the upsampling showed that putting the movie review data at the end achieved the best scores for three out of the seven datasets. The performances overall were superior to the scores of the downsampling. Using the movie data as the last dataset in the training resulted in a $78.18\%$ macro f1-score for the scientific data which is $1.21\%$ better compared to setting the scientific data at the end of the sequence. The downsampled part shows that the training with the product data, in the end, has shown the best performance for the three datasets. Interestingly, the performance on the scientific data was $8.21\%$ better using the downsampled either the product or movie datasets in the end compared to using its own dataset as last in the downsampled scenario. Except for the testing on Instruments and the CSC-Clean dataset, the performances of the other datasets did not change dramatically based on the feeding sequence. Another interesting finding was that putting the movie reviews in the end for the downsampled experiments did not result in a bad performance for all other dataset categories and led to a maximum drop of $2.86\%$ for the US Airline dataset compared to the best performance for that dataset. In general, it was not the case that the models shows a bias towards the dataset that was used last in the training epoch. It is to be noted that due to the computational effort we did not try every combination but selected a subset that puts every category once at each position. Furthermore, the general finding of this experiment series is that unexpectedly the network does not work better when trained last on the evaluating dataset. Although most of the achieved accuracies are comparable it is not easy to predict which sequence works best for which testing set. Generally, upsampling was superior for most of the datasets. However, it requires much more training time. In our case, the dataset size is $3,000$ compared to $797$ samples for the downsampled version.

\subsubsection{Experiment 3: Shuffled Training.}
In addition to the sequential data feeding experiment, we performed similar experiments by shuffling the data. The major difference compared to the previous experiment was that there is no sequence preserved, neither within the categories nor between the categories. Therefore, the gradients can align to each of the data samples and are not biased towards the last category in the setup.

\subsubsection{Results and Discussion:}
In Table~\ref{tab:shuffled} we show the results of the shuffled upsampling and downsampling experiments. Surprisingly, the macro-f1 scores are close to each other. In these experiments, the downsampled data used about $800$ instances of each dataset whereas the upsampled $3,000$. Even more interesting is that the shuffled model performed well across all datasets. The largest accuracy drop compared to the single dataset training models was about $3.44\%$ for the Sentiment140 dataset. Comparing the performances of the downsampled model to the models trained exclusively on those datasets, the accuracy of the shuffled model is impressively good. The same holds for the upsampled model. In general, the shuffled model holds a better generalization as it can be applied on all the datasets even without fine-tuning and sticks to good performance.

\begin{table}[!t]
\renewcommand{\arraystretch}{1.3}
\caption{Macro f1-scores for shuffled training. 'Up' corresponds to the upsampled training data and 'Down' to the downsampled training data.}
\label{tab:shuffled}
\centering
\scalebox{0.96}{
\begin{tabular}{l||c|c|c||c||c|c||c}
\multirow{2}{*}{\diagbox[width=4.5em]{\textbf{Train}}{\textbf{Test}}}
& \multicolumn{3}{c||}{\textbf{Movie}} & \textbf{Product} & \multicolumn{2}{c||}{\textbf{Twitter}} & \textbf{Scientific} \\
\cline{2-8}
& IMDB & Cornell & Stanford & Instruments & Us Airline & Sentiment140 & CSC-Clean \\
\hline\hline
Up      & $93.65$ & $88.04$ & $91.81$ & $88.90$ & $89.99$ & $77.13$ & $74.45$ \\
\hline\hline
Down    & $97.80$ & $87.07$ & $88.42$ & $83.40$ & $86.96$ & $76.65$ & $73.73$ \\
\end{tabular}
}
\end{table}

\subsection{Multi-task Model: Fusing Scientific Sentiment and Intent}

\subsubsection{Experiment 1: Multi-Domain Usage}
We further experimented with the unified model for the sentiment and intent classification. This experiment combines both tasks into a single model. The motivation behind this experiment is to handle the increased amount of computation resource and inference time when using two separated models as proposed in ImpactCite. However, due to the small size of the CSC-Clean dataset it is not possible to train it directly in conjunction with the intent task. Therefore, we utilized the previous findings and combined the citation sentiment data with the sentiment datasets from other domains to enlarge the training set. Therefore, the sentiment task covers the sentiment classification for all used datasets that included a neutral class.

\subsubsection{Results and Discussion:}

\begin{table}[!t]
\renewcommand{\arraystretch}{1.3}
\caption{Macro f1-scores sentiment and intent classification. Shows that the single task model is superior for the individual tasks.}
\label{tab:multi-task}
\centering
\begin{tabular}{l||c|c||c|c}
\multirow{2}{*}{\diagbox[width=5.5em]{\textbf{Task}}{\textbf{Setup}}}
& \multicolumn{2}{c||}{\textbf{Mutli-Task}} & \multicolumn{2}{c}{\textbf{Single-Task (ImpactCite)}} \\
\cline{2-5}
& All sent. datasets & CSC-Clean $+$ Stanford & CSC-Clean & SciCite \\
\hline\hline
Sentiment   & $64.00$ & $56.00$ & $\bm{80.41}$ & $*$ \\
Intent      & $78.00$ & $78.00$ & $*$ & $\bm{88.93}$ \\
\end{tabular}
\end{table}

Results in Table~\ref{tab:multi-task} show that the unified multi-task model has advantages however it is achieved with certain limitations. Firstly, the advantage of the multi-task model is that only a single model is used and two different heads are trained.
This makes inference twice as fast as only one forward pass is needed and reduces the required hardware. However, the only impediment is that the model is trained on the conjunction of sentiment data and therefore the bias of the out-domain context can hinder the intent performance. It is to be noted that the model is robust against out-domain data for the sentiment task.

\section{Discussion}
In our previous paper~\cite{mercier2020impactcite} we have shown that our approach is capable to perform well on both the sentiment and intent classification. The results clearly highlighted the problems with the scientific sentiment domain and the lack of data. Additionally, the unbalanced datasets resulted in difficulties to converge for all evaluated methods except ImpactCite~\cite{mercier2020impactcite}. Neither ALBERT~\cite{lan2019albert} nor BERT~\cite{devlin2018bert} were able to converge up to a state that provides a sufficient performance across all tested classes. While an intent classification using those models works well this is not the case for sentiment classification as some classes were not captured by the models. Especially, the negative class was identified as one of the major shortcomings. However, we were able to overcome this data shortcoming up to a certain extent using ImpactCite~\cite{mercier2019senticite}. We achieved a new state-of-the-art performance for both tasks emphasizing the gains using XLNet~\cite{yang2019xlnet} when the existing data is limited and unbalanced. In addition, these findings served as a baseline for qualitative citation analysis which is most times not considered due to the lack of available datasets.

In this paper we mainly focused on the utilization of out-domain data to enhance the sentiment classification in the scientific domain which suffers from the lack of existing annotated datasets. Our experiments have shown that without a specific fine-tuning the correlation between in-domain datasets is stronger compared to out-domain datasets and it is possible to achieve surprisingly good results training a classifier on a dataset of the same domain even without fine-tuning. Interestingly, in some cases, the larger quality datasets have shown better performance on some test sets than using the original training set. Going one step ahead, we evaluated different scheduling techniques to better understand the impact of data fusion. First, we tried different sequential concatenations resulting in better-generalized models that we are able to perform well across all datasets. Although the sequence has been shown to bias the performance slightly towards the last category the results showed that the movie data as the last set in the sequence performed best. In addition, the difference between the upsampled and downsampled training dataset versions highlighted that if the number of datasets concatenated is sufficient then this approach works for very small datasets below $800$ samples. Next, we mixed all sentiment training data to avoid preserving sequence to favor any of the domains which resulted in a superior model with respect to the generalization. Shuffling all the data removed the convergence towards a single domain. Although it would be possible to fine-tune the model on a single dataset. We demonstrate that our solution is more robust as it is confronted with out-domain data during the training and further utilizes this data to establish a more general understanding of the underlying language concepts that are not bound towards one domain.

Ultimately, the combination of tasks within a single model can be very complex. During our experiments, we faced several challenges while combining the sentiment and intent tasks. It was not possible to train a model that is capable to converge using only the scientific sentiment and intent data. This is the case as the sentiment data is very small and when combined with the intent task, the network is not able to learn the concept of sentiment, especially negative sentiment, due to a large amount of unrelated data. Although we have shown in our previous work~\cite{mercier2020impactcite} that the use of two separate models is possible this might not be desired as the hardware required to run two models parallel is expensive. Furthermore, a sequential inference suffers from time delay. As a feasibility study, we combined the sentiment data with the out-domain sentiment data and trained the multi-task model. Ultimately, the proposed model is capturing multiple tasks and domains.

\section{Conclusion}
Utilizing our previous conducted experiments and findings presented in~\cite{mercier2020impactcite} we evaluated the impact of out-domain data usage during the training to enhance datasets and overcome data scarcity in less popular domains. Specifically, the issues faced in the sentiment analysis motivated us to evaluate the combination of different domain sentiment tasks and our results show impressive performances when the training procedure is aligned to work with the multiple concatenated datasets. Our first finding highlights that training using an in-domain dataset can already result in a suitable classifier for the target dataset even without fine-tuning due to the correlation of the data within the same domain. Going one step further we evaluated the impact of mixed datasets across the domains to enlarge the available amount of data. Doing so we found that the results for some datasets could be improved using a sequential approach in which the datasets with higher quality at the end boost the classifier. Furthermore, shuffling the datasets resulted in a powerful cross-domain model showing a good performance across all datasets. In contrast to the sequential scheduling, the performance of the shuffled approach was more balanced and not biased towards a single domain. Ultimately, we have shown in a feasibility study that multi-task models can be enhanced using out-domain data to enlarge the dataset. It was impossible to combine the scientific sentiment and citation data directly using the sentiment data due to the scarcity of the data. However, with out-domain data mixing, and a shuffled schedule we were able to come up with a fully converge sentiment and intent model. One benefit of this model is the shared encoder resulting in much lower hardware requirements, faster training, and inference. In contrast to that, the separately trained models better converge for their specific task resulting in higher accuracies. We aim for the optimization of the dataset combinations and task combinations to achieve a better multi-task model open for future research.



\bibliographystyle{splncs04}
\bibliography{bibliography}

\end{document}